\documentstyle[twocolumn,floats,aps,epsf]{revtex}

\begin{document}

\twocolumn[\hsize\textwidth\columnwidth\hsize\csname@twocolumnfalse%
\endcsname

\title{How universal is the one-particle Green's function of a 
Luttinger liquid?}
\author{V.\ Meden}
\address{Insitut f\"ur Theoretische Physik C, Technische Hochschule
Aachen, D-52056 Aachen, Germany}

\date{September 30, 1998}
\maketitle
\draft
\begin{abstract}
The one-particle Green's function 
of the Tomonaga-Luttinger model for one-dimensional
interacting Fermions is discussed.
Far away from the origin of the plane of space-time 
coordinates the function falls off like a power law. 
The exponent depends on the direction within the plane.  
For a certain form of the interaction
potential or within an approximated cut-off procedure the 
different exponents 
only depend on the strength of the interaction at zero momentum
and can be expressed in terms of the Luttinger liquid parameters 
$K_{\rho}$ and $K_{\sigma}$ of the model at hand.
For a more general interaction and directions which are determined by the 
charge velocity $v_{\rho}$ and spin velocity $v_{\sigma}$ the exponents also 
depend on the ``smoothness'' of the interaction at zero momentum
and the asymptotic behavior of the Green's function is not given 
by the Luttinger liquid parameters alone. This shows that the 
physics of large space-time distances in Luttinger liquids 
is less universal than is widely believed.  
\end{abstract}
\pacs{PACS numbers: 71.10.-w, 71.10.Pm}
]

\narrowtext

\smallskip

Over the past 20 years it has been shown that the low energy physics 
of a variety of models of one-dimensional (1D) correlated electrons
can be described by the Luttinger liquid (LL) 
phenomenology\cite{haldane,johannes}. Because of the 
progress in the experimental realization of quasi 1D systems and
speculations about possible LL behavior in the normal state of the 
high-temperature superconductors LL phenomenology
has lately attracted considerable attention. 
In Fermi
liquids the low energy physics and low temperature thermodynamics 
is dominated by the excitation of quasi-particles
which are in a one-to-one correspondence to the particle-hole
excitations of the non-interacting system. The elementary excitations
in LL's are of collective bosonic nature. In LL's the 
discontinuity of the momentum distribution function $n(k)$ at the Fermi
wave vector $k_F$ vanishes and the density of states of the
one-particle Green's function at the Fermi
energy is zero. Close to the Fermi surface
both functions are dominated by power law behavior with exponents
which are given by the so called LL parameters $K_{\rho}$ and $K_{\sigma}$.  
According to LL phenomenology thermodynamic quantities 
(e.\ g.\ the compressibility and spin susceptibility) 
and the non-analytic behavior
of correlation functions can be expressed in terms of 
these two parameters and the velocities
$v_{\rho}$ and $v_{\sigma}$ of charge and spin excitations. 
By calculating the one-particle Green's
function of the Tomonaga-Luttinger (TL)
model\cite{tomonaga,luttinger} we will show that the asymptotic 
behavior of the Green's function of a general LL 
at large space-time distances is nonetheless less universal than is 
widely believed. The exponents of the algebraic decay in the
directions within the $x$-$v_F t$ plane which are 
determined by the velocities $ v_{\rho}$ and $v_{\sigma}$ 
are {\it not} given by the LL 
parameters of the 
model alone. Here $v_F$ denotes the Fermi velocity. 
We will furthermore discuss the implications for the
spectral function.

The TL  model is a continuum model of interacting
1D electrons.
The linearization of the electron dispersion around the two Fermi
points and neglecting the backscattering processes between
electrons
makes it feasible to determine the spectrum of the
Hamiltonian and calculate all correlation functions. 
Following Luttinger\cite{luttinger} we introduce right- ($\alpha=+$) 
and left-moving ($\alpha=-$)
Fermions with spin $s$, creation operators $a^{\dag}_{k,\alpha,s}$, 
dispersion $\xi_{\alpha}(k)= \alpha v_F (k- \alpha k_F)$,
density operators ($q \neq 0$) $\rho_{\alpha,s}(q)= \sum_k 
a_{k,\alpha,s}^{\dagger} a_{k+q,\alpha,s}^{}$, and particle
number operators $n_{k,\alpha,s}=a_{k,\alpha,s}^{\dagger} 
a_{k,\alpha,s}^{}$. To simplify the mathematical treatment Luttinger
added an infinite filled Fermi sea to the
ground state\cite{luttinger}. 
The Hamiltonian for a system of length $L$ is given by
\begin{eqnarray}
&& H  =  \sum_k \sum_{\alpha,s} \xi_{\alpha}(k) 
\left[ n_{k,\alpha,s} - \left< n_{k,\alpha,s}  \right>_0 \right] \nonumber \\
& & + \frac{1}{2 L} \sum_{{q \neq 0} \atop {\alpha, s, s'}} 
\left[g_{4,\parallel}(q) \delta_{s,s'} + 
g_{4,\perp}(q) \delta_{s,-s'} \right]
\rho_{\alpha,s}(q) \rho^{\dag}_{\alpha,s'}(q) \nonumber \\
&&  +   \frac{1}{L} \sum_{{q \neq 0} \atop {s, s'}} 
\left[g_{2,\parallel}(q) \delta_{s,s'} + 
g_{2,\perp}(q) \delta_{s,-s'} \right]
\rho_{+,s}(q) \rho^{\dag}_{-,s'}(q) . 
\label{hamiltonian}
\end{eqnarray}
Here we use ``g-ology'' notation\cite{solyom} and
$\left< \ldots \right>_0$ denotes  the 
(non-interacting) ground state expectation value.
Contrary to many authors we keep the explicit $q$  
dependence of the coupling functions. We assume that the Fourier 
transforms $g_{i,\kappa}(q)$ ($i=2,4$; $\kappa=\parallel , \perp$)
of the two-particle interaction have only contributions 
for $q {< \atop \sim} q_c \ll k_F$ 
with an interaction cut-off $q_c$. At {\it no} stage of the discussion 
it will be necessary to introduce any further cut-offs 
``by hand'' despite the infinite (filled) 
Fermi sea at negative energies. The model only belongs to the LL
universality class if $g_{2,\kappa}(q=0)$ is finite for 
$\kappa = \parallel$ and  $\perp$ and at least
one of the two coupling constants is non-zero. 
Thus we restrict ourselves to these kind of 
interactions. 

Bosonization of the Hamiltonian and a canonical transformation
leads to\cite{johannes}
\begin{eqnarray}
H = \sum_{q \neq 0} \sum_{\nu = \rho,\sigma} \varepsilon_{\nu}(q) 
\beta_{\nu}^{\dagger}(q) \beta_{\nu}^{}(q) ,  
\label{hamiltoniandiag} 
\end{eqnarray}
with bosonic operators $\beta_{\nu}(q)^{\dagger}$ describing 
charge ($\nu=\rho$) and spin ($\nu=\sigma$) excitations 
(spin-charge separation).
The energies $\varepsilon_{q,\nu}$ are given by 
\begin{eqnarray} 
\frac{\varepsilon_{q,\nu}}{|q|}   =  v_F  \sqrt{\left( 1+
\frac{g_{4,\nu}(q)}{\pi v_F} \right)^2 -
\left( \frac{g_{2,\nu}(q)}{\pi v_F}\right)^2  } 
 \equiv   v_{\nu}(q)  ,
\label{energies} 
\end{eqnarray}
where we have introduced the renormalized charge and spin density
velocities $v_{\nu}(q)$ and interactions 
$g_{i,\rho/\sigma}(q) \equiv \left[g_{i,\parallel}(q) 
\pm g_{i,\perp}(q) \right]/2$.
The one-particle Green's function $i G_{\alpha,s}^{<} (x,t)  = 
 \left< \psi_{\alpha,s}^{\dagger}(0,0) 
\psi_{\alpha,s}^{}(x,t)\right>$,
which after a double Fourier transformation 
leads to the spectral function $\rho_{\alpha,s}^{<}(k,\omega)$
relevant for photoemission experi\-ments, can be calculated using the
bosonization of the fermion fields
$\psi_{\alpha,s}^{\dagger}(x)= \frac{1}{\sqrt{L}} \sum_{k} e^{-ikx} 
a_{k,\alpha,s}^{\dagger}$\cite{haldane}. In the
thermodynamic limit and at zero temperature we obtain 
$ G_{+}^{<}(x,t)  = [G_{+}^{<}]^0(x,t) \exp{\{F(x,t)\}}$
with 
\begin{eqnarray}
F(x,t) & = & \frac{1}{2}\sum_{\nu=\rho, \sigma} 
\int_{0}^{\infty} \frac{dq}{q} \left\{
e^{- i q \left( x - v_{\nu}(q)  t \right) } 
- e^{- i q \left( x -  v_F  t \right) } \right. \nonumber \\* && + 
\left. 2 \gamma_{\nu}(q) \left[ \cos{(qx)} e^{i q v_{\nu}(q) t} -1 \right]  
\right\} 
\label{fdef} 
\end{eqnarray}
and the non-interacting Green's function
\begin{eqnarray}
[G_{+}^{<}]^0(x,t) 
= - \frac{1}{2\pi} \frac{e^{i k_F x} }{  x-v_F t -i0  }.
\label{g0t0}
\end{eqnarray} 
Because the Green's function is the same for
both spin directions we have suppressed the spin index $s$.
By leaving out irrelevant particle number contributions in the 
Hamiltonian we effectively shifted the energy scale so that the 
chemical potential is zero.  
$\gamma_{\nu}(q)$ in Eq.\ (\ref{fdef}) is given by the eigenvectors of
the canonical transformation and can be written as
$\gamma_{\nu}(q)=\left[ K_{\nu}(q)+ 1/K_{\nu}(q) -2 \right]/4$, 
with 
\begin{eqnarray}
K_{\nu} (q)= \sqrt{\frac{1+g_{4,\nu}(q)/(\pi v_F)-g_{2,\nu}(q)/(\pi
    v_F)}{1+g_{4,\nu}(q)/(\pi v_F)+g_{2,\nu}(q)/(\pi
    v_F)}} .
\label{kdef}
\end{eqnarray}
Due to the interaction cut-off the momentum integral in Eq.\ (\ref{fdef}) 
is regular in the ultraviolet limit. 
The $q \to 0$ limits of the four functions $v_{\nu}(q)$ and $K_{\nu} (q)$ 
define the two velocities and two LL parameters 
($v_{\rho}, v_{\sigma}, K_{\rho}, K_{\sigma}$). 

Roughly speaking the non-analyticities in spectral functions at low energies
and small momenta are given by the behavior of $G_{+}^{<}(x,t)$ at
large arguments. To determine the non-analytic
behavior of the momentum 
distribution  $n_+(k) = \int_{-\infty}^{\infty} dx e^{-ikx} i G_+^<(x,0)$
and the momentum {\it integrated} spectral
function $\rho_+^<(\omega)= \int_{-\infty}^{\infty} dt  e^{i \omega t}
 i   G_+^<(0,t)/(2 \pi)$ it is sufficient to discuss the behavior of
$G_{+}^{<}(x,t)$ along the $x$ and $v_F t$ axis. To emphasize the 
difference to the general case of non-vanishing $x$ {\it and} $v_F t$ 
we first focus on these two cases. With a little less
mathematical rigor they have already been discussed several
times\cite{approxg}. 
Using integration by parts\cite{regular} in the first derivative 
of $F(x,0)$ and 
$F(0,t)$ we obtain the {\it leading} behavior 
$F(x,0) \sim - \alpha \ln{ \left| x \right|}$ and 
$F(0,t) \sim - \alpha \ln{ \left| v_F t \right|}$, with $\alpha = 
\gamma_{\rho}(0) + \gamma_{\sigma}(0)$. In $G_{+}^{<}(x,t)$ 
this leads to power law behavior along 
the $x$ and $v_F t$ axis 
with an exponent which {\it only} depends on the strength of the 
interaction at vanishing momentum and thus the LL parameter $K_{\rho}$ 
and $K_{\sigma}$ of the model at hand. This is in accordance with 
the LL phenomenology. 
From general theorems about Fourier transforms 
the by now well known LL behavior of the momentum distribution
function follows for $|k-k_F|/q_c \to 0$ 
\begin{eqnarray*}
\frac{1}{2} - n_+(k) \sim \left\{ 
\begin{array}{r@{\quad:\quad}l}
|k-k_F|^{\alpha} \mbox{sign} (k-k_F) & \mbox{for} \,\, 0 <\alpha <1 \\
(k-k_F) \ln{|k-k_F|} & \mbox{for} \,\, \alpha =1 \\
(k-k_F) & \mbox{for} \,\, \alpha > 1 .    
\end{array} \right. 
\end{eqnarray*}
\vspace{-0.6cm}
\begin{eqnarray}
\label{nk}
\end{eqnarray}
For $\alpha > 1$ the leading behavior of $n_+(k)$ is dominated 
by a linear term but a higher derivative still diverges at $k_F$.    
Due to the different analytic properties of $F(0,t)$ (analytic in the
upper half of the complex $t$ plane) and $F(x,0)$ ({\it not} analytic in
either the upper or the lower half of the complex $x$ plane) 
$\rho_+^<(\omega)$ is dominated by a
non-analytic power law behavior even for $\alpha > 1$\cite{vmdoc}. 
Thus anomalous dimensions $\alpha >1$ should in principle be 
observable in momentum integrated photoemission spectra\cite{experiments}. 
In the literature the mathematical reason for this important
difference in the behavior of $n_+(k)$ and 
$\rho_+^<(\omega)$ has thus far not been properly pointed out.      
For $\omega/(v_F q_c) \to 0$ we obtain 
\begin{eqnarray}
\rho_+^<(\omega) \sim  \left\{ 
\begin{array}{r@{\quad:\quad}l}
\Theta(-\omega) (-\omega)^{\alpha} & \mbox{for} \,\, \alpha \not\in 
{\rm I \! N} \\
\Theta(-\omega) (-\omega)^{\alpha} \ln{|\omega|} & \mbox{for} 
\,\, \alpha \in {\rm I \! N}  .  
\end{array} \right. 
\label{rhoomega}
\end{eqnarray}
The prefactors and the range over which the leading
behavior given in  
Eqs.\ (\ref{nk}) and (\ref{rhoomega}) can be observed depend on
the interaction at {\it all} $q$. As a consequence numerically calculated
curves for different interaction potentials $g_{i,\kappa}(q)$ but the
same anomalous dimension might appear quite 
different\cite{vmdoc,ksvm}.  

To find a more explicit form of the Green's function 
the $q$ integral in Eq.\ (\ref{fdef}) has often been evaluated 
after replacing $v_{\nu}(q) \to v_{\nu}(0)$, $\gamma_{\nu}(q) \to 
\gamma_{\nu}(0)$ and multiplying the integrand by a factor 
$\exp{\left(-q\Lambda\right)}$\cite{approxg,vmks}. 
The $q$ integral can then be performed and one obtains
\begin{eqnarray}
&& [G_{+}^{<}]_A (x,t)   =    \frac{-e^{ik_F x}/(2 \pi)}
{x- v_F  t -i0} \prod_{\nu =\rho,\sigma} 
\left[ \frac{x- v_F  t - i \Lambda}
{x- v_{\nu} t - i \Lambda } \right]^{1/2} \nonumber \\* 
&& \times \left[ \frac{\Lambda^2}
{\left( x- v_{\nu}  t - i \Lambda \right)
 \left( x+ v_{\nu}  t + i \Lambda \right) } 
\right]^{\gamma_{\nu}/2} .
\label{gapprox}
\end{eqnarray}   
For the Hamiltonian Eq.\ (\ref{hamiltonian}) there exists 
{\it no} special interaction potential so that this
{\it approximation} becomes exact.
As in the discussion of the general case
below we will transform onto new variables $s=x-ct$ and $s'=x+ct$ with
an arbitrary velocity $c$ and discuss the behavior of $ [G_{+}^{<}]_A \left( 
x[s,s'],t[s,s']\right)$ for large $s$ with a fixed $s'$ and vice versa
and for different values of $c$. For all velocities $c$ but $v_{\rho}$ and
$v_{\sigma}$, $ [G_{+}^{<}]_A \left( x[s,s'],t[s,s']\right)$ falls off
like $s^{-(1+\alpha)}$ and $s'^{-(1+\alpha)}$. This is the behavior we
already found for $v_F t =0$ or $x=0$. 
For $c=v_{\rho}$ the Green's function falls off like 
\begin{eqnarray} 
[G_{+}^{<}]_A \left( x[s,s'],t[s,s'] \right) 
& \sim & s^{-(1 + \gamma_{\sigma} + \gamma_{\rho}/2)} , 
\label{gabfallvfs1} \\ 
\lbrack G_{+}^{<} \rbrack_A \left( x[s,s'],t[s,s'] \right) 
& \sim & s'^{-(1/2+ \gamma_{\sigma} +\gamma_{\rho}/2)} .
\label{gabfallvfss1} 
\end{eqnarray}
The behavior of the Green's function for $c=v_{\sigma}$ follows from 
Eqs.\ (\ref{gabfallvfs1}) and (\ref{gabfallvfss1}) 
by interchanging $\gamma_{\rho}$ and $\gamma_{\sigma}$.
Within the above approximation the exponents of the 
asymptotic behavior for all $c$ and thus for all directions within the
$x$-$v_F t$ plane can be expressed in terms of 
$K_{\rho}$ and $K_{\sigma}$.
The resulting spectral function 
displays power law singularities (bounded non-analyticities 
in case the $\gamma_{\nu}$ are too large) at 
$\omega = \pm v_{\nu} (k-k_F)$ for all 
$\Lambda (-k+k_F)>0$\cite{vmks}.  

Without using any approximations it has been shown that 
Eq.\ (\ref{gapprox}) gives the correct
asymptotic behavior of $G_{+}^{<} \left( x,t\right)$ for 
a {\it box potential} 
$g_{i,\kappa}(q) = g_{i,\kappa} \Theta(q_c - |q|)$\cite{ksvm}.
In this case the momentum range over which $\rho_{+}^{<}(k,\omega)$ 
is dominated by two power law singularities at the charge and spin
excitation energies is limited to $0<(-k+k_F)/q_c < 1$. For other
momenta further non-analyticities occur\cite{ksvm}.

Next we will show that $[G_{+}^{<}]_A (x,t)$ does {\it not} 
give the correct
asymptotic behavior for an arbitrary shape of the interaction.  
After transforming onto $s$ and $s'$  the Green's function is given by
\begin{eqnarray}
G_{+}^{<} \left( x[s,s'],t[s,s'] \right)  =
[G_{+}^{<}]^0 \left( x[s,s'],t[s,s'] \right)   e^{\tilde{F}(s,s')} ,
\label{gt0sss}
\end{eqnarray} 
where $\tilde{F}(s,s')$ follows from  Eq.\ (\ref{fdef}) and 
$(x,t) \to (s,s')$.
As an example we will discuss the behavior of $\tilde{F}(s,s')$ 
for large $s'$ and fixed
$s$ in more detail and only present the results for the other case. 
We first take the derivative with respect to $s'$. This gives
\begin{eqnarray*}
&& \frac{d\tilde{F}(s,s')}{ds'}  =  \frac{1}{2}\sum_{\nu=\rho,\sigma} 
\int_{0}^{\infty} 
dq \left\{ - i  \frac{1}{2} \left( 1- \frac{ v_{\nu}(q)}{c}\right)
\nonumber \right. \\ 
&& \times e^{- i q \frac{1}{2} \left( 1+ \frac{ v_{\nu}(q)}{c}\right)  s }
e^{- i q \frac{1}{2} \left( 1- \frac{ v_{\nu}(q)}{c}\right)  s' }  
\nonumber  \\   
&& + i  \frac{1}{2} \left( 1- \frac{v_F}{c}\right)
e^{- i q \frac{1}{2} \left( 1+ \frac{ v_F}{c}\right)  s }
 e^{- i q \frac{1}{2} \left( 1- \frac{ v_F}{c}\right)  s' }  \nonumber 
\\ && + \gamma_{\nu}(q) \left[  
i  \frac{1}{2} \left( 1+ \frac{ v_{\nu}(q)}{c}\right)
e^{ i q \frac{1}{2} \left( 1- \frac{ v_{\nu}(q)}{c}\right)  s }    
e^{ i q \frac{1}{2} \left( 1+ \frac{ v_{\nu}(q)}{c}\right)  s' }    
\nonumber \right. \\
&& \left. \left.  - i  \frac{1}{2} 
\left( 1- \frac{ v_{\nu}(q)}{c}\right)
e^{- i q \frac{1}{2} \left( 1+ \frac{ v_{\nu}(q)}{c}\right)  s }    
e^{- i q \frac{1}{2} \left( 1- \frac{ v_{\nu}(q)}{c}\right)  s' }   
 \right]   \right\}  .
\end{eqnarray*}
\vspace{-0.6cm}
\begin{eqnarray}
\label{ftildeabgel}
\end{eqnarray}   
To simplify the discussion we assume
that $ v_{\nu}(q)$ are monotonic functions. 
Using integration by parts and the method of stationary phase in the
asymptotic expansion of the integral we find for 
all velocities $c$ but $v_{\rho}(0)$, $v_{\sigma}(0)$, and $v_F$ that 
$\tilde{F}(s,s')$ goes like $- \alpha \ln{(s')}$
and thus
\begin{eqnarray}
\tilde{G}_{+}^{<} \left( s,s' \right) \equiv
G_{+}^{<} \left( x[s,s'],t[s,s'] \right) \sim s'^{
  -(1+\alpha) } .
\label{gallg} 
\end{eqnarray}
For large $s$ and fixed $s'$ we obtain the same behavior.

If $c=v_{\rho}(0)$ the phase $q \left[ 1-v_{\rho}(q)/v_{\rho}(0)
\right]$ in the first and fourth term 
of Eq.\ (\ref{ftildeabgel}) becomes stationary at $q=0$. 
At the stationary point also the
prefactor $\left[ 1-v_{\rho}(q)/v_{\rho}(0) \right]$ vanishes and thus
we have to {\it generalize} the method of stationary phase. 
The details of this generalization will be given elsewhere
and here we will only present the results. 
The leading contribution of
the fourth term of Eq.\ (\ref{ftildeabgel}) to the large $s'$ behavior
is $- \frac{\gamma_{\rho}}{2} \frac{1}{p_{\rho} +1} \frac{1}{s'}$, where 
$p_{\rho}$ is the smallest number $n \in {\rm I \! N} \cup \{ \infty \}$ 
with $\left[v_{\rho} \right]^{(n)}(0) \neq 0$, where 
$\left[v_{\rho} \right]^{(n)}(q)$ denotes the $n$-th derivative.  
According to the definition of $v_{\rho} (q)$ 
Eq.\ (\ref{energies}) $p_{\rho}$ is a measure of the ``smoothness'' of the
interaction at vanishing momentum. Evaluating the first term in Eq.\
(\ref{ftildeabgel}) in the same way and using integration by parts in
the other terms leads to
\begin{eqnarray}  
 \tilde{G}_{+}^{<} \left( s,s' \right)  
\sim  s'^{-(1/2+ \gamma_{\sigma} +\gamma_{\rho}/2 
+ 1/[2p_{\rho}+2]+ \gamma_{\rho}/ [2 p_{\rho}+2])}.
\label{gabfallvfss1allg} 
\end{eqnarray}
This result is different from Eq.\ (\ref{gabfallvfss1}) and more
importantly it shows that the asymptotic behavior {\it cannot} be 
obtained from the LL parameters $K_{\rho}$ and 
$K_{\sigma}$ of the TL model alone. This is in {\it contrast} 
to the wide spread belief that Eq.\ (\ref{gapprox}) displays the 
asymptotic behavior of {\it all} models which belong to the LL 
universality class provided the exponents are expressed in terms of
the LL parameters of the specific model. The behavior Eq.\
(\ref{gabfallvfss1}) is only recovered if $p_{\rho}=\infty$, i.\ e.\
if all derivatives of $v_{\rho} (q)$ and thus all derivatives of the 
interaction potential vanish at $q=0$.
For this reason Eq.\ (\ref{gapprox}) gives the correct leading
behavior in case of a box potential.   
For the large $s$ behavior and $c=v_{\rho}(0)$ we obtain
\begin{eqnarray} 
 \tilde{G}_{+}^{<} \left( s,s' \right) 
 \sim  s^{-(1+ \gamma_{\sigma} +\gamma_{\rho}/2 
+ \gamma_{\rho}/ [2 p_{\rho}+2])}.
\label{gabfallvfs1allg} 
\end{eqnarray}
For $c=v_{\sigma}(0)$ we have to analyze the Green's function following
the same route and obtain  
\begin{eqnarray}  
 \tilde{G}_{+}^{<} \left( s,s' \right)  
& \sim & s^{-(1 + \gamma_{\rho} + \gamma_{\sigma}/2
+ \gamma_{\sigma}/ [2 p_{\sigma}+2])}
,
\label{gabfallvfs2allg} \\
 \tilde{G}_{+}^{<} \left( s,s' \right) 
& \sim & s'^{-(1/2+ \gamma_{\rho} +\gamma_{\sigma}/2
+ 1/[2p_{\sigma}+2]+ \gamma_{\sigma}/ [2 p_{\sigma}+2])} ,
\label{gabfallvfss2allg} 
\end{eqnarray}
where $p_{\sigma}$ is defined in analogy to $p_{\rho}$.

The case $c=v_F$ has so far been excluded. If $v_F \neq v_{\rho}(0)$
and $v_F \neq v_{\sigma}(0)$ the behavior of the Green's function 
is given by Eq.\ (\ref{gallg}). In the relevant case of a spin
independent interaction with $v_{\sigma}(q) \equiv v_F$ and $\gamma_{\sigma}
=0$ the Green's function is given by Eqs.\ (\ref{gabfallvfs2allg}) and
(\ref{gabfallvfss2allg}) with $p_{\sigma} = \infty$.   

\begin{figure}[htb]
\begin{center}
\vspace{-1.0cm}
\leavevmode
\epsfxsize \columnwidth
\epsffile{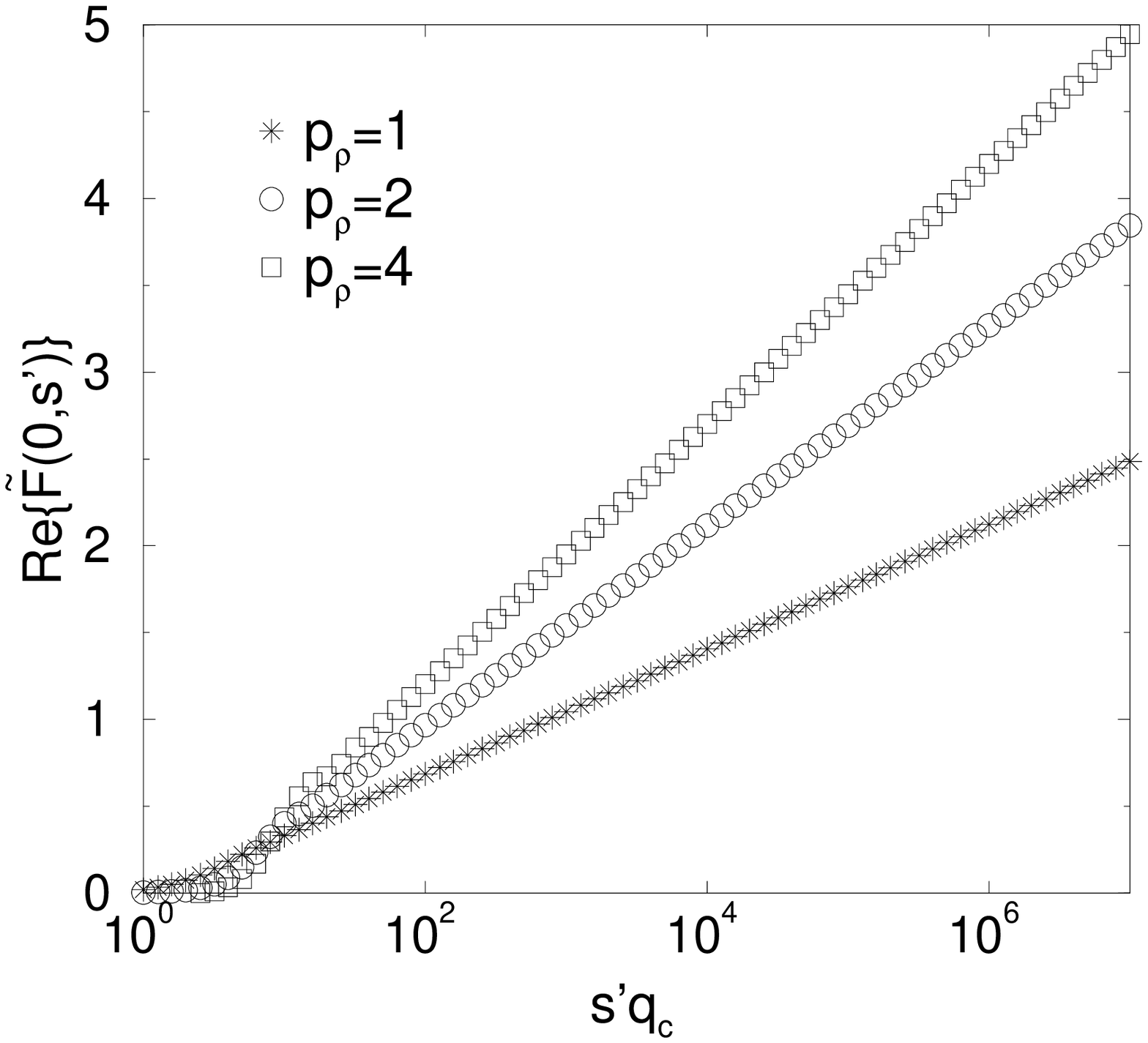}
\caption{$\mbox{Re} \, \left\{
\tilde{F}(0,s') \right\}$ as a function of $s'q_c$ for different 
$p_{\rho}$. For details see the text.}
\label{fig1}
\epsfxsize \columnwidth
\epsffile{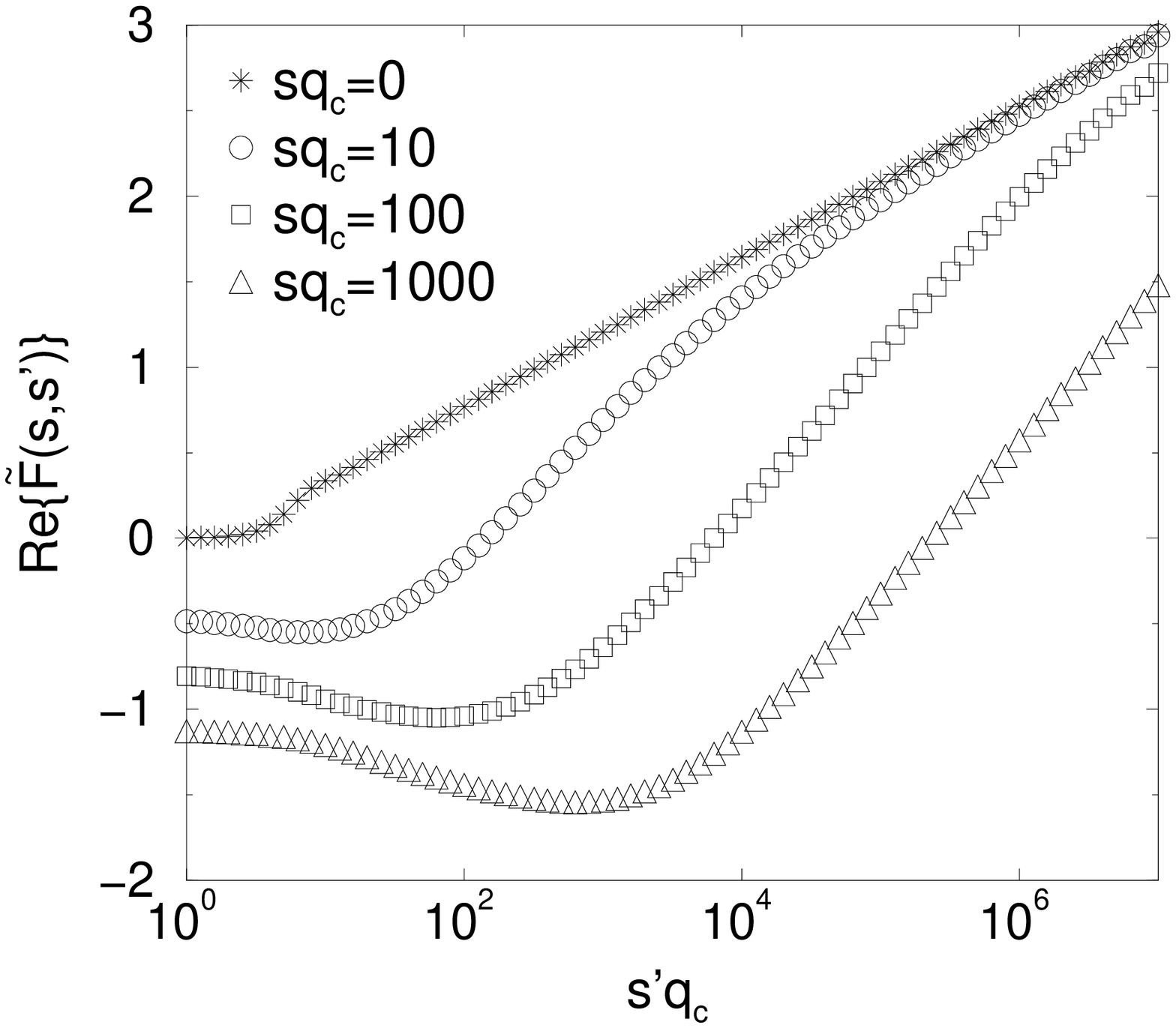}
\caption{$\mbox{Re} \, \left\{
\tilde{F}(s,s') \right\}$ as a function of $s'q_c$ for different
$s q_c$. For details see the text.}
\label{fig2}
\end{center}
\vspace{-0.5cm}
\end{figure}

To gain more insight into the asymptotic behavior of the Green's
function we have evaluated the momentum integral in 
$\tilde{F}(s,s')$ numerically. As an illustration of the 
new predictions Eqs.\
(\ref{gabfallvfss1allg})-(\ref{gabfallvfss2allg}) 
we present $\mbox{Re} \, \{
\tilde{F}(s,s') \}$ for $s=0$ as a function of $s'$ and
$c=v_{\rho}(0)$ on a log-linear scale in Fig.\ \ref{fig1}. 
The curves have been calculated for a spin independent interaction
with
$g_{4,\parallel}(q) \equiv g_{4,\perp}(q) \equiv 
 g_{2,\parallel}(q) \equiv  g_{2,\perp}(q) \equiv g \exp{ \{
 ( -q/q_c)^{p_{\rho}} \} }$, $2 g/(\pi v_F) = 3$, i.\ e.\ 
$v_{\rho}(0)/v_F =2$, $\gamma_{\rho} =1/8$ and different $p_{\rho}$.  
From Eqs.\ (\ref{gt0sss}) and (\ref{gabfallvfss1allg}) we expect   
\begin{eqnarray}
\tilde{F}(s,s') \sim \left\{ \frac{1}{2} -  
\left( \frac{\gamma_{\rho}}{2} + \frac{1}{2p_{\rho}+2}
+ \frac{\gamma_{\rho}}{2 p_{\rho}+2} \right)
\right\} \ln{\left(s' \right)}
\label{expect}
\end{eqnarray}
Fits of the data for $s' > 10^3$ reproduce the expected prefactors of 
the $\ln{(s')}$ within a relative error of less than 0.02\%. 
For increasing 
$s$ we have to go to larger $s'$ to find the asymptotic behavior 
Eq.\ (\ref{expect}). In an intermediate regime of $s'$, which 
increases with increasing $s$,  
$\mbox{Re} \, \{ \tilde{F}(s,s') \}$ 
displays the behavior of the $p_{\rho} = \infty$ case. This is
illustrated in Fig.\ \ref{fig2}. It shows 
$\mbox{Re} \, \{ \tilde{F}(s,s') \}$
as a function of $s'$ on a log-linear scale for the above form of 
coupling functions, $c=v_{\rho}(0)$ and different $s$. 
The parameters are $2 g/(\pi v_F) = 5$, i.\ e.\ 
$v_{\rho}(0)/v_F =\sqrt{6}$, $\gamma_{\rho} =0.2144$ and
$p_{\rho} =2$.  
A fit for $s q_c=0$ again reproduces the expected behavior Eq.\ (\ref{expect}) 
with a very high accuracy. The data for $s q_c=10$ only show this behavior
for $s'q_c > 10^5$. For $s q_c=1000$ a fit for $10^6<s' q_c< 10^7$ gives 
$0.3934 \ln{(s')}$ which is very close to $0.3929 \ln{(s')}$, the
expected behavior for $p_{\rho} = \infty$. Similar to the  
$s q_c=100$ curve for arguments between $10^6$ and $10^7$, 
the curve for $s q_c=1000$ shows a cross over 
at very large $s' q_c$ and the prefactor of the logarithmic term is 
again given by the $p_{\rho} = 2$ value $0.1904$.

For the approximated Green's function Eq.\ (\ref{gapprox})
and the box potential the two-dimensional Fourier transformation 
which leads to the momentum resolved spectral 
function
can be performed analytically\cite{vmks,ksvm}. 
Thus far we have not succeeded in calculating 
$\rho_{\alpha,s}^{<}(k,\omega)$ for an interaction with $p_{\nu} <
\infty$.
From the Fourier transformation of the approximated
Green's function it is known, that the exponents of the
algebraic decay along the special directions ($c=v_{\nu}$)  
determine the exponents of the 
non-analyticities at $\omega = \pm v_{\nu} (k-k_F)$. As we have shown 
above the exponents of the algebraic decay
of $G_{+}^{<} ( x,t)$ along the special directions are different from the 
exponents of $[G_{+}^{<}]_A (x,t) $, thus we have {\it no} reason
to believe that $\rho_{+}^{<}(k,\omega)$ shows power law singularities
with the same exponents as 
$[\rho_{+}^{<}]_A(k,\omega)$. 
{\it It is not even obvious that  the exact spectral function shows 
power law singularities at all.} 
Certainly we expect 
the two peak structure of spin and charge excitations, but it is not
clear if, for any non-vanishing $k-k_F$, these peaks are given by
algebraic singularities. The extended region in which 
$G_{+}^{<} ( x,t)$ displays
the asymptotic behavior of $[G_{+}^{<}]_A (x,t) $ on the other hand 
indicates that the resulting spectral functions might
look very similar at least for very small $|k-k_F|$. For 
$|k-k_F| \to 0$ we expect to find growing regions in which the exact 
spectral function resembles the power law behavior of the 
approximation (with the same exponents as the approximation) up to 
energies very close to 
$\pm v_{\nu} (k-k_F)$ but {\it not} exactly at these energies.
A comparison of broadened spectral 
functions for a finite system, i.\ e.\ of spectral functions 
with no ``real'' algebraic singularities, for a box and Gaussian 
potential is presented in Ref.\ \cite{vmdoc} and indeed shows a prominent 
similarity between both spectra. A more detailed comparison
will be given elsewhere. 

The results presented here are important for the interpretation of    
numerically calculated Green's functions and spectra of microscopic 
models.  
They show that it is impossible to determine the LL 
parameters of the considered model from the asymptotic behavior of the
Green's function within the special directions. The numerical
evaluation of the asymptotic behavior of the Green's function gives
on the other hand a possibility to confirm our predictions.
Furthermore it should be possible to confirm the predictions
by analyzing the finite size scaling of the weight of single peaks 
in the spectral function of a microscopic model of finite length. 
For energies close to but different from $\pm v_{\nu} (k-k_F)$ 
we expect to find power law behavior with exponents given by the 
LL paramters as in the approximation. Not so for the scaling 
of the peaks exactly at $\pm v_{\nu} (k-k_F)$.
Our results have also 
consequences for the comparison of angle resolved and angle integrated
spectral functions which have been measured by high resolution 
photoemission\cite{experiments}. 


The author would like to thank K.\ Sch\"onhammer, W.\ Metzner, and N.\
Shannon for very helpful discussions.


\begin{thebibliography}{}
\bibitem{regular} We assume that the interaction is sufficiently
  regular.
\end{thebibliography}
\end{document}